\documentclass[twocolumn, 10pt, pra, aps, showpacs, reprint, amsmath, amssymb, superscriptaddress, nofootinbib]{revtex4-1}

\usepackage[]{graphicx}
\usepackage{color}

\newcommand{\pbyp}[2]{\frac{\partial #1}{\partial #2}}
\DeclareMathOperator{\sinc}{\mathrm{sinc}}

\DeclareMathOperator{\ez}{\vec{e}_z}
\newcommand{\I}{\mathrm{I}}
\newcommand{\II}{\mathrm{II}}
\newcommand{\III}{\mathrm{III}}
\DeclareMathOperator{\psisup}{\psi_\mathrm{sup}}
\renewcommand{\Re}{\mathrm{Re}}

\begin{document}

\title{Dynamical control of square microlaser emission via symmetry classes}
\author{S. Bittner}
\author{A. Loirette-Pelous}
\author{C. Lafargue}
\affiliation{Laboratoire de Photonique Quantique et Mol{\'e}culaire, UMR 8537, Ecole Normale Sup{\'e}rieure de Paris-Saclay, CentraleSup{\'e}lec, CNRS, Universit{\'e} Paris-Saclay, 94235 Cachan, France}
\author{I. Gozhyk}
\affiliation{Surface du Verre et Interfaces (SVI), UMR 125 CNRS/Saint-Gobain Recherche, 93303 Aubervilliers, France}
\author{C. Ulysse}
\affiliation{Centre de Nanosciences et de Nanotechnologies, CNRS, Univ. Paris-Sud, Universit\'e Paris-Saclay, C2N Marcoussis, 91460 Marcoussis, France}
\author{B. Dietz}
\affiliation{School of Physical Science and Technology, and Key Laboratory for Magnetism and Magnetic Materials of MOE, Lanzhou University, Lanzhou, Gansu 730000, China}
\author{J. Zyss}
\author{M. Lebental}
\email{melanie.lebental@ens-cachan.fr}
\affiliation{Laboratoire de Photonique Quantique et Mol{\'e}culaire, UMR 8537, Ecole Normale Sup{\'e}rieure de Paris-Saclay, CentraleSup{\'e}lec, CNRS, Universit{\'e} Paris-Saclay, 94235 Cachan, France}

\date{\today}

\begin{abstract}
A major objective in photonics is to tailor the emission properties of microcavities which is usually achieved with specific cavity shapes. Yet, the dynamical change of the emission properties during operation would often be advantageous. The implementation of such a method is still a challenging issue. We present an effective procedure for the dynamical control of the emission lobes which relies on the selection of a specific coherent superposition of degenerate modes belonging to different symmetry classes. It is generally applicable to systems exhibiting pairs of degenerate modes. We explored it experimentally and analytically with organic square microlasers, which emit narrow lobes parallel to their sidewalls. By means of the pump polarization, emission lobes are switched on and off selectively with an extinction ratio better \mbox{than $1/50$}.
\end{abstract}

\pacs{42.55.Sa, 03.65.Sq, 05.45.Mt, 42.60.Da}

\maketitle

\section{Introduction}
Microcavities have been widely studied because of the high sensitivity of their emission to their shape. Accordingly, they provide appealing solutions for numerous applications, from on-chip photonic devices \cite{matsko-livre} to bio-chemical sensing \cite{revue-capteur-bio,revue-capteur-chimie}. While circular microdisks exhibit an isotropic in-plane emission, the cavity geometry can be deformed in numerous ways to favor emission in a given direction, starting from an ellipse \cite{schwefel} with integrable ray dynamics, to mixed-dynamics systems with a shape from the Lima\c{c}on family \cite{limacon} to the chaotic stadium \cite{lebental-stade-PRA}. Since the far-field pattern results from the resonator geometry, it can normally be modified only by fabricating other devices. Accordingly, the dynamical control of resonator properties has become an intense field of research. Some
proposals are still theoretical \cite{pola-aleatoire,JAP-theorie-QD,srep-spintransistor}, whereas others have been applied successfully in experiments \cite{tahar, sebbah, huicao-disk, JMGerard, srep-polariton-Berry, PRL-phonons, APL-hall-optique}. 

We present a procedure to dynamically control the directions of emission by selecting specific coherent superpositions of degenerate modes via the pump polarization. This has the advantage of not requiring a spatial light modulator as in Refs.~\cite{tahar, sebbah, huicao-disk, srep-polariton-Berry}. Its efficacy is demonstrated with organic square microlasers, though it is applicable to various photonic systems and beyond. The underlying principle is indeed of general validity, since it is based on fundamental symmetry considerations and exploits the sensitivity to external influences of systems exhibiting degenerate modes. 

\begin{figure}[tb]
\begin{center}
\includegraphics[width = 1\linewidth]{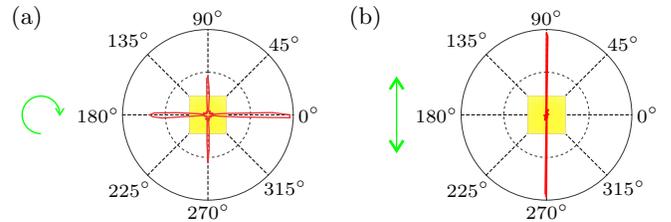}
\end{center}
\caption{Experimental far-field intensity distributions of a square microlaser. A photograph of the cavity is superimposed to indicate the orientation. (a) Circular pump polarization, side length $a=120\,\mu$m. (b) Linear pump polarization, side length $a=180\,\mu$m~. Due to technical reasons, the accessible angles for Fig.~\ref{fig:FFexp}(b) were limited to [$-15^{\circ},195^\circ$]; the full polar plot was created by duplicating the data.}
\label{fig:FFexp}
\end{figure}

In contrast to the isotropic emission of disks or spheres, organic square-shaped microlasers emit four narrow lobes \cite{EPL-carre}. This specific feature is well described by a semiclassical model~\cite{huang-carre, EPL-carre, Bittner2013}. A comprehensive review on square microcavities is provided in Ref.~\cite{huang-carre}. An important characteristic of this approach is the twofold degeneracy of all resonant modes suggesting that each lasing state is composed of a linear superposition of two modes and is consequently sensitive to any exterior perturbation. We used the pump polarization state as an exterior parameter and observed a spectacular manifestation of the coherent superposition of degenerate modes: in the far-field emission diagram, some lobes are simply switched on and off. Under circular pump polarization, four lobes are observed parallel to the sidewalls [Fig.~\ref{fig:FFexp}(a)], consistent with the symmetries of a square, whereas, if the symmetry is broken by a linear pump polarization, only two lobes remain parallel to the polarization direction [Fig.~\ref{fig:FFexp}(b)]. 

The outline of the article is the following. First, the general concepts are explained in Sec.~\ref{sec:mainideas} and validated experimentally in Sec.~\ref{sec:exp}. Then, a comprehensive theoretical description explains the observations and reveals the universally applicable concepts in Sec.~\ref{sec:theo}. 

\begin{figure}[tb]
\begin{center}
\includegraphics[width = 1\linewidth]{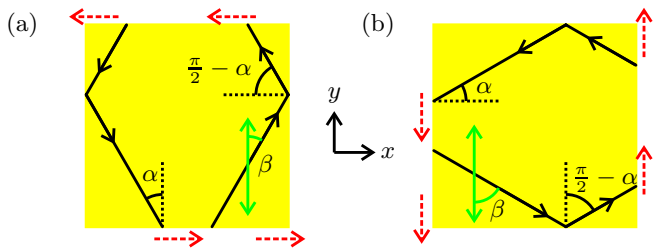}
\end{center}
\caption{(Color online) Schematic drawings of wave vectors (black solid lines) inside a square cavity. Each wave function in Table 1 is composed of 8 plane waves. Their corresponding wave vectors belong to two different subsets presented in (a) and (b) where (b) is obtained from (a) by a $\pi/2$ rotation. Each subset is dynamically closed, i.e., a ray remains in its subset after reflections at the sidewalls. If the angle of incidence $\alpha$ is close to the critical angle ($\alpha_c \simeq 42^{\circ}$) emission takes place at a grazing angle while maintaining good confinement. The complementary angle of incidence $\pi/2-\alpha$ is significantly larger than $\alpha_c$. Thus the ray is confined inside the cavity by total internal reflection, leading to negligible emission. Consequently, emission (red dashed arrows) occurs along the $x$ direction in (a) and along the $y$ direction in (b). The pump polarization is indicated by a green doublesided arrow. The pump efficiency depends on the angle $\beta$. It is maximal for a plane wave with $\beta = \pi/2$. In the case of a pump polarization along the $y$ direction, $\beta$ is slightly larger in (b) than in (a), and hence configuration (b) is enhanced by stimulated emission.}
\label{fig:rayGeom}
\end{figure}

\section{Basic ideas} \label{sec:mainideas}
An organic square microlaser emits four narrow lobes parallel to its sidewalls [see Fig.~\ref{fig:FFexp}(a)], because the light is emitted at a grazing angle. Within the semiclassical approach, the corresponding resonant modes are well described by $8$ plane waves \cite{EPL-carre}, which can be inferred from one another by reflection at the boundary and a $\pi/2$ rotation, see Fig.~\ref{fig:rayGeom}. 

\begin{table*}[t]
\caption{Symmetry classes, quantum numbers, and semiclassical wave functions. The symbols '$+$' or '$-$' mean that the wave function is even or odd, respectively, with respect to the symmetry line. $\psi_0$ is the normalization factor. Here, the wave functions are normalized such that the integral of $\vert\psi(x,y)\vert^2$ over the cavity area equals unity. The Mulliken symbols are defined in Ref.~\cite{tinkham}. Reprinted from Ref.~\cite{EPL-carre}.}
\label{tab:dlmWFs}
\vspace{3 mm}
\begin{center}
\begin{tabular}{cccccl}
\hline
\hline
Diagonal & Horizontal/vertical & Parity of & Parity of & Mulliken & Model wave function \\
symmetry & symmetry & $m_x + m_y$ & $m_x \cdot m_y$ & symbol & \\
\hline
$(++)$ & $+$ & Even & Even & $A_1$ & $\psi(x, y) = \psi_0 [\cos(k_x x) \cos(k_y y) + \cos(k_y x) \cos(k_x y)]$ \\
$(--)$ & $+$ & Even & Even & $B_2$ & $\psi(x, y) = \psi_0 [\cos(k_x x) \cos(k_y y) - \cos(k_y x) \cos(k_x y)]$ \\
\hline
$(++)$ & $-$ & Even & Odd & $B_1$ & $\psi(x, y) = \psi_0 [\sin(k_x x) \sin(k_y y) + \sin(k_y x) \sin(k_x y)]$ \\
$(--)$ & $-$ & Even & Odd & $A_2$ & $\psi(x, y) = \psi_0 [\sin(k_x x) \sin(k_y y) - \sin(k_y x) \sin(k_x y)]$ \\
\hline
$(+-)$ & None & Odd & Even & $E$ & $\psi(x, y) = \psi_0 [\sin(k_x x) \cos(k_y y) + \cos(k_y x) \sin(k_x y)]$ \\
$(-+)$ & None & Odd & Even & $E$ & $\psi(x, y) = \psi_0 [\sin(k_x x) \cos(k_y y) - \cos(k_y x) \sin(k_x y)]$ \\
\hline
\hline
\end{tabular}
\end{center}
\end{table*}

Accordingly, all resonances can be assigned to pairs of degenerate modes\footnote{The semiclassical model features three pairs of degenerate mode classes. See Sec.~\ref{sec:theo} for a detailed discussion of mode degeneracy.}, see Table 1. The wave functions in a pair, e.g., $\psi_{A_{1}}$ and  $\psi_{B_{2}}$, feature a far-field emission pattern similar to the experimental one with four lobes in the directions parallel to the sidewalls, see Fig.~\ref{fig:FFexp}(a). However, a coherent superposition of $\psi_{A_{1}}$ and  $\psi_{B_{2}}$ --- which is possible for a given wavelength due to their degeneracy --- does not necessarily exhibit the same symmetric far-field pattern. We demonstrate that the polarization state of the pump laser can select either the superposition $\psi_+ = \psi_{A_{1}} + \psi_{B_{2}}$ or $\psi_- = \psi_{A_{1}} - \psi_{B_{2}}$, which corresponds to selecting four out of the eight plane waves, that is, configuration (a) or (b) in Fig.~\ref{fig:rayGeom}. The control by the pump polarization is achieved via the commercial laser dye DCM\footnote{4-(Dicyanomethylene)-2-methyl-6-(4-dimethylaminostyryl)-4Hpyran, provided by Exciton.}, for which absorption and emission transition dipoles are parallel. They are isotropically and uniformly distributed within the cavity. 

Under circular pump polarization, all the dyes are excited without symmetry breaking, and the far-field emission patterns thus feature four lobes, in accordance with the global symmetry of the square. Under a linear pump polarization, the laser dyes which are parallel to this direction are preferentially excited. They then emit the well-known $\sin^2(\theta)$ dipole diagram (where $\theta$ is the angle with the dipole axis), which is maximal in the direction perpendicular to the pump polarization. For instance, in the case of a pump polarization parallel to the $y$ axis, emission in the $x$ direction is favored, which corresponds to $\beta = \pi/2$ in Fig.~\ref{fig:rayGeom}. So plane waves along the trajectories shown in Fig.~\ref{fig:rayGeom}(b) are more likely to be excited than along those in Fig.~\ref{fig:rayGeom}(a), and the laser amplification selects the state of which the wave vectors are more aligned with the $x$ direction. As sketched in Fig.~\ref{fig:rayGeom}(b), this state and the associated plane waves are confined by total internal reflection on the sidewalls parallel to the $x$ direction, whereas their incidence is close to the critical angle on the sidewalls parallel to the $y$ axis, thus yielding an emission along the $y$ direction. Consequently, due to the mediation by the cavity resonances,  in the case of the square the laser emission is parallel to the pump polarization, not perpendicular to it as one would expect for dye molecules in free space. 

\begin{figure}[tb]
\begin{center}
\includegraphics[width = 6.0 cm]{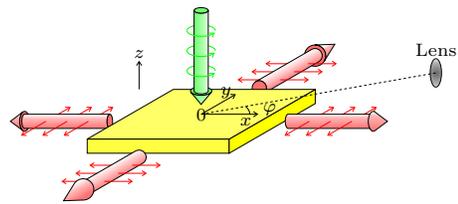}
\end{center}
\caption{(Color online) Geometry of the setup. The pump beam (vertical thick green arrow) impinges vertically on the cavity. The microlaser emits in the cavity plane (horizontal thick red arrows). The emission is collected by a lens in the far field at azimuthal angle $\varphi$. The thin arrows indicate the polarization states.}
\label{fig:setupGeom}
\end{figure}

\section{Experimental setup and results} \label{sec:exp}
The microlaser cavities are fabricated from a PMMA matrix\footnote{Poly(methyl methacrylate), PMMA A6 resist provided by Microchem.} doped with $5$ wt\% of the laser dye DCM. The 650~nm-thick polymer cavities are etched by electron beam lithography on a Si/SiO$_2$ ($2~\mu$m) substrate. This allows to fabricate sidewalls and corners with excellent quality at a nanoscopic scale \cite{EPL-carre}. The side lengths of the square cavities vary from $a = 50~\mu$m to $200~\mu$m. The pump beam from a frequency-doubled Nd:YAG laser ($532$~nm, $0.5$~ns, $10$~Hz) impinges vertically and with uniform intensity onto a single cavity (see Fig.~\ref{fig:setupGeom}). Its polarization is controlled using half- and quarter-waveplates. All experiments were performed at room temperature. 

In the experiments with a circularly polarized pump beam, the collection lens in the far field was fixed and the microlaser rotated around its own axis \cite{OE-3D}. In the case of a linearly polarized pump beam, the cavity was kept fixed and the collection lens was moved around the cavity with a 3D goniometer \cite{APL-FP}. Then the angular range of observation was reduced to $[-15^{\circ}, 195^{\circ}]$ for technical reasons, but the direction of pump polarization remained aligned with the cavity with a precision better than 1$^{\circ}$. 

\begin{figure}[tb]
\includegraphics[width = 1\linewidth]{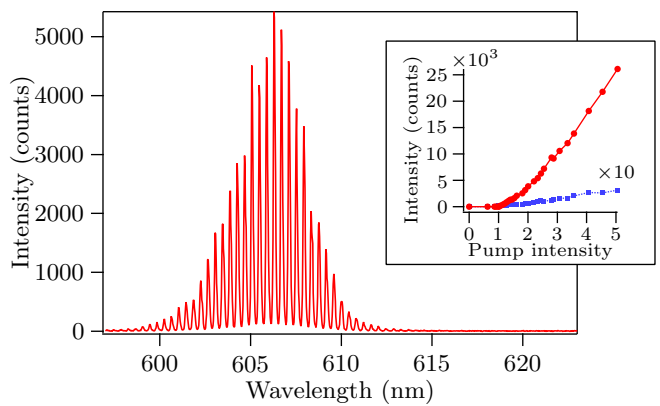}
\caption{Experimental spectrum of a $200~\mu$m large square microlaser. The pump is linearly polarized parallel to one pair of sidewalls, and the observation direction is parallel to the pump polarization. The pump intensity is about twice the lasing threshold. Inset: threshold curve for pump polarization parallel (solid red line) and perpendicular (dotted blue line) to the direction of observation. The maximum intensity of the spectrum is plotted versus the pump intensity, which is normalized to the laser threshold at $3.0$~MW$\cdot$cm$^{-2}$. }
\label{fig:seuil}
\end{figure}

The square microlasers emit a comb-like multi-mode spectrum \cite{EPL-carre, PRA-spectres}, with a clear lasing threshold (see Fig.~\ref{fig:seuil}). The far-field intensity distribution under circular pump polarization is plotted in Fig.~\ref{fig:FFexp}(a), exhibiting four emission lobes in line with the symmetry axes of the square. For linear pump polarization, the symmetry of the system is effectively reduced as illustrated in Fig.~\ref{fig:FFexp}(b). The ratio of the intensity in the $\pm x$ directions over the intensity in the $\pm y$ directions is less than $2 \cdot 10^{-2}$. Note that the speed of the dynamical control is only limited by the time needed to change the pump polarization. The microlaser emission itself is predominantly polarized parallel to the cavity plane. 

\section{The model} \label{sec:theo}
In this section, the basic ideas and observations outlined above are corroborated with model calculations applicable to any polarization of the pump beam. The main quantity of interest is the probability $\mathcal{P}$ that a mode is supplied by the photons emitted by the laser dyes. We assume that the experimentally observed lasing modes are the ones with the lowest lasing threshold, which have the highest probability $\mathcal{P}$. Since we restrict ourselves to the situation near the lasing threshold, we may base our model on passive resonances without mode interaction. First, the semiclassical model developed in \cite{EPL-carre} is briefly sketched. Then the derivation of the probability $\mathcal{P}$ is explained step-by-step. Finally, we determine the superposition $\psi_\textrm{sup}$ of, e.g., wave functions $\psi_{A_1}$ and $\psi_{B_{2}}$ from the first group in Table~\ref{tab:dlmWFs}, for which the probability $\mathcal{P}$ of exciton transfer to $\psi_\textrm{sup}$ is maximized. 

The square microlasers we are studying can be assumed two-dimensional \cite{PRA-spectres}. They emit predominantly in the transverse electric (TE) polarization state with the electric field in the cavity plane. Thus, they are governed by the two-dimensional Helmholtz equation $\left( \Delta + n^2 k^2 \right) \psi\, = 0$ for the $z$-component of the magnetic field, $\psi = B_z$, with appropriate boundary conditions, where $k$ denotes the modulus of the wave vector $\vec k$ and $n \simeq 1.5$ the effective refractive index. The Helmholtz equation for a metallic square cavity can be solved analytically for Dirichlet and  for Neumann boundary conditions. On the contrary, for the dielectric square this is not possible due to diffraction at the dielectric corners~\cite{gennarelli-2015}. Therefore, we proposed a semiclassical model in Ref.~\cite{EPL-carre} which describes the salient experimental features and also agrees well with numerical simulations. 

In the following we briefly summarize the semiclassical model. The wavenumber components of $\vec k=(k_x,k_y)$ are quantized by the boundary conditions, which lead to two quantum numbers $m_{x,y}$. A wave function inside the cavity is composed of a plane wave propagating with $\vec k(m_x,m_y)=(k_x,k_y)$ as well as its reflections at the boundary and rotations by $\pi/2$. These plane waves may be expressed in terms of the model wave functions listed in Table 1. According to the $C_{4v}$ symmetry of the square, the wave functions belong to six different symmetry classes, namely $A_1$, $A_2$, $B_1$, $B_2$, and $E$ \cite{tinkham, McIsaac}. The dimension of the sets $A$ and $B$ is 1 and it is 2 for $E$. The character of the $A$ set is 1, while it is -1 for the $B$ set. Since the parities of $m_x + m_y$ and of $m_x m_y$ coincide for each mode pair ($A_1$, $B_2$), ($A_2$, $B_1$), and ($E$, $E$), the semiclassical quantization condition yields the same $\vec{k}$ and hence each mode is twofold degenerate. 

However, this degeneracy is broken by the diffraction at the dielectric corners, which is not taken into account by the model. Thus the actual modes $\psi_{A_1}$ and $\psi_{B_2}$, as well as $\psi_{A_2}$ and $\psi_{B_1}$, are not degenerate. Full numerical calculations reproduce the expected tiny lifting of the degeneracy of the resonances of these two pairs of symmetry classes, but also demonstrate that they are degenerate in very good approximation and very well described by the semiclassical model. Especially for realistic cavity sizes, this effect is negligible since the relative effect of the corners becomes smaller and smaller with increasing cavity size. In contrast, the modes of the two $E$ symmetry classes are exactly degenerate because the square belongs to the $C_{4v}$ point symmetry group. Hence we can treat all three mode pairs as exactly degenerate in the following. 

\begin{figure}[tb]
\begin{center}
\includegraphics[width = 1\linewidth]{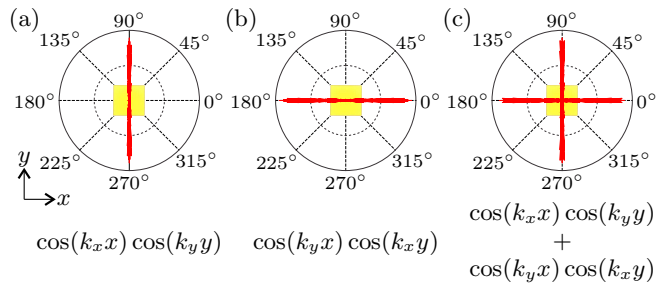}
\end{center}
\caption{Far-field intensity distributions calculated analytically with the Kirchhoff integral (see Appendix \ref{AppI}), based on different superpositions of $\psi_{A_1}$ and $\psi_{B_2}$ as indicated below each figure. We used $(m_x, m_y) = (444, 398)$, which corresponds to ($k_x = 1.16259 \cdot 10^5 ~\mathrm{cm}^{-1}$, $k_y = 1.04336 \cdot 10^5~\mathrm{cm}^{-1}$). (a) $\psi_+=\psi_{A1}+\psi_{B2}$. (b) $\psi_- = \psi_{A_1} - \psi_{B_2}$. (c) $\psi_{A_1}$.}
\label{fig:calcul-far-field}
\end{figure}

In Ref.~\cite{EPL-carre}, we evidenced that the observed lasing modes correspond to an angle of incidence just above the critical angle, $\alpha_c\simeq 42^{\circ}$, which is consistent with grazing emission parallel to each sidewall. Let us consider for instance the two-dimensional subset ($A_1$, $B_2$). The lasing wave function is then either $\psi_{A_{1}}$ or $\psi_{B_{2}}$, or a linear superposition of them. For a given pair ($m_x$, $m_y$) with $m_x>m_y$ and thus $\mathrm{Re}(k_x)>\mathrm{Re}(k_y)$, $\psi_{A_{1}} + \psi_{B_{2}}$ emits two lobes along the $y$ direction and none along the $x$ direction [see Fig.~\ref{fig:calcul-far-field}(a) and Fig.~\ref{fig:rayGeom}(b)], while it is the opposite for $\psi_{A_{1}} - \psi_{B_{2}}$ [see Fig.~\ref{fig:calcul-far-field}(b) and Fig.~\ref{fig:rayGeom}(a)]. Since the other two mode pairs exhibit the same basic structure of the far-field distributions and selection rules for the four emission lobes as a function of the pump polarization, we only discuss the exemplary case of ($A_1$, $B_2$) in the following. 

Which superposition is selected for lasing is determined by the interaction with the active medium and the pump polarization. A dye-doped lasing medium can exhibit gain anisotropy, i.e., the gain experienced by a plane wave depends on its propagation direction \cite{valeur,lakowicz}. This is a consequence of the structure of the dye molecules. Their absorption and emission processes can be described by the corresponding dipole transition moments $\vec{d}_a$ and $\vec{d}_e$, respectively \cite{lakowicz, valeur, iryna-polarisation, Gozhyk2015}. For a DCM molecule, the pump laser excites an electronic $S_0 \rightarrow S_1$ transition, and the absorption dipole is oriented mostly along the axis of the molecule \cite{iryna-polarisation}. After a fast non-radiative relaxation in the vibrational manifold of the $S_1$ state, spontaneous or stimulated emission occurs from the $S_1$ to the $S_0$ state. For many dyes, including DCM, the emission dipole from the $S_1$ to the $S_0$ state is parallel to the absorption dipole \cite{iryna-polarisation}, so the effect described here is of broad validity. In the following, we therefore assume that the emission and absorption dipoles are exactly parallel and denote them by $\vec d$. 

The unit vector describing the polarization state of the pump beam is denoted by $\vec p=(p_x,p_y)$. Then, for a given dye with a dipole moment $\vec d$, the excitation cross section is proportional to $|\vec d\cdot\vec p\,|^2$. Furthermore the probability that the dye transfers its excitation to a mode with a given electric field $\vec E$ is proportional to $|\vec d\cdot\vec E(x,y)\,|^2$ for a dye located at the position $(x,y)$. The laser dyes are assumed to be uniformly and isotropically distributed. Therefore, the probability $\mathcal{P}$ to excite the mode is
\begin{equation} \label{eq:probabilite-integrale}
\mathcal{P} \propto \int_S \mathrm{d}S \int_{\Omega} \mathrm{d}\Omega \left| \vec d(\Omega)\cdot\vec E (x,y) \right|^2 \left| \vec d(\Omega)\cdot\vec{p} \right|^2
\end{equation}
where the integrals are over the surface $S$ of the cavity and the orientations $\Omega$ of the dipoles. Among a set of modes with similarly high $Q$-factors, those with the highest $\mathcal{P}$ have the lowest threshold. 

We outline in the following how $\mathcal{P}$ was calculated for the square microlasers. Since the microlaser emission is polarized in-plane we only consider TE polarized modes. In this case the wave functions $\psi$ correspond to the magnetic field component $B_z$. The electric field is therefore given by $\vec{E} \propto\vec{e}_z \times \vec \nabla B_z \propto \left(-{\partial \psi}/{\partial y}, {\partial \psi}/{\partial x}, 0 \right)$. Inserting this expression into Eq.~(\ref{eq:probabilite-integrale}) and integrating over $\Omega$ yields
\begin{equation}
\label{eq:integralesurA} \begin{array}{c} \mathcal{P}\propto d^4 \int_S \mathrm{d}S \bigg[ (2 |p_y|^2 + 1) \left| \frac{\partial \psi}{\partial x} \right|^2 \\ + (2 |p_x|^2 + 1) \left| \frac{\partial \psi}{\partial y} \right|^2 - 4 \mathrm{Re}{(\frac{\partial \psi}{\partial y} \frac{\partial \psi^*}{\partial x})} \mathrm{Re}{(p_x p_y^*)} \bigg] \, . \end{array} \end{equation}
The last step consists in calculating the superposition of wave functions which maximizes $\mathcal{P}$ for a given pump polarization. The whole space of possible superpositions can be covered by two real parameters $\gamma\in[0,\pi)$ and $\delta\in[-\pi/2,\pi/2)$, yielding for the pair ($A_1, B_2$)
\begin{equation} \label{eq:superposition}
\psi_\textrm{sup} = \cos(\gamma) \psi_{A_{1}} + \mathrm{e}^{i \delta} \sin(\gamma) \psi_{B_{2}} \, .
\end{equation}
The calculations are tedious but straightforward and are summarized in Appendix~\ref{AppII}. It appears that ${\partial P}/{\partial\delta} \propto \sin\delta$ and therefore the probability $\mathcal{P}$ is maximal for $\delta=0$ independently of the pump polarization. In contrast, the optimal $\gamma$ depends on the pump polarization. If it is along the $y$ direction, $\vec p = (0, 1)$, then $\gamma_\mathrm{max} = \pi/4$, that is,
\begin{equation}
\psi_\textrm{sup}(x, y) \propto \cos(k_x x) \cos(k_y y)=\psi_+ \, ,
\end{equation}
and the emission is parallel to the $y$ direction. Similarly, if the pump polarization is along the $x$ direction $\vec p=(1,0)$, then $\gamma_\mathrm{max} = 3 \pi / 4$ yielding
\begin{equation}
\psi_\textrm{sup}(x, y) \propto \cos(k_y x) \cos(k_x y)=\psi_- \, ,
\end{equation}
and the emission is parallel to the $x$ direction. Under a circular pump polarization $\vec p\propto (1, \pm\,i)$ we obtain $\gamma_\mathrm{max} = 0$, so $\psi_\textrm{sup}$ is equal to $\psi_{A_{1}}$, which features four equal emission lobes in the $x$ and $y$ directions. These values of $\gamma_\mathrm{max}$ are the same for the two other degenerate mode pairs ($A_2, B_1$) and ($E, E$), leading to the same experimental far-field patterns for all lasing modes. 

\section{Conclusions}
We demonstrated experimentally that different superpositions of degenerate modes can be excited in a square organic microlaser via the pump polarization and can be discerned from their vastly different far-field emission patterns. These observations were predicted based on a semiclassical model within the theoretical framework of fluorescence anisotropy. While the proposed method is demonstrated for an organic square microlaser, it does not depend on its specific geometry nor on the dye-doped polymer as gain medium and thus is of broad applicability. The only requirements are the existence of degenerate mode pairs with distinct target features (e.g., emission direction) and of a procedure to select a specific superposition of them. 

Degenerate mode pairs exist for all resonators with symmetry classes $C_{Nv}$ where $N \geq 3$, and microlasers with such symmetries like equilateral triangles or hexagons are commonplace \cite{Braun2000, Nobis2005, Li2006, Yoon2007, Yang2007, Lafargue2014}. Furthermore, so-called degenerate cavities can exhibit highly degenerate modes \cite{Arnaud1969, Nixon2013, Pascal2017}. While the selection of a superposition via the pump polarization is particularly simple, it is restricted to dye-based lasers. However, other selection schemes like spatially modulated pumping via an SLM or segmented metal contacts can be used for, e.g., semiconductor lasers. Accordingly, this principle of dynamic control via degenerate modes is bound to find applications in many microphotonic systems and beyond. 

\begin{acknowledgments}
S.~B.\ gratefully acknowledges funding from the European Union Seventh Framework Programme (FP7/2007-2013) under Grant No.\ 246.556.10. This work was supported by a public grant from the Laboratoire d'Excellence Physics Atom Light Matter (LabEx PALM) overseen by the French National Research Agency (ANR) as part of the Investissements d'Avenir program (Reference No.\ ANR-10-LABX-0039). This work was supported by the French RENATECH network (French national nanofabrication platform). ENS Paris-Saclay (invited professorship to B.~D.) is acknowledged for supporting the collaboration. B.~D.\ thanks the NSF of China for financial support under Grant No. 11775100. The authors thank Marc Guillon for careful reading of the manuscript.
\end{acknowledgments}

\appendix

\section{Calculation of the Far-Field Distribution} \label{AppI}
We use the Kirchhoff integral,
\begin{equation} \psi(\vec{r}) = \oint_{\partial S} \, \mathrm{d}|\vec{r}{\,'}| \left\{ \, G(\vec{r}, \vec{r}{\,'}) \pbyp{\psi}{n}(\vec{r}{\,'}) \right. \left. - \pbyp{G}{n}(\vec{r}, \vec{r}{\,'}) \psi(\vec{r}{\,'}) \right\} \, , \end{equation}
to determine the far-field distribution at a location $\vec r$ outside the domain $S$ covered by the square resonator \cite{jackson}, which is assumed to be two-dimensional. Here, $\pbyp{}{n}$ is the derivative in the direction of the outward normal to the resonator boundary $\partial S$ and
\begin{equation} G(\vec{r}, \vec{r}{\,'}) = \frac{H_0^{(1)}(k |\vec{r} - \vec{r}{\,'}|)}{4 i} \end{equation}
is the free-space Green's function in two dimensions, where $k = (k_x^2 + k_y^2)^{1/2} / n$ is the wave number in free space, and $k_x,\, k_y$ are the wave vector components inside the resonator with refractive index $n$. We impose at $\partial S$ dielectric boundary conditions on the wave functions $\psi_\mathrm{in}$ and $\psi_\mathrm{out}$ inside and outside the domain, respectively, $\psi_\mathrm{in}\vert_{\partial S}=\psi_\mathrm{out}\vert_{\partial S}$ and $\mu\pbyp{\psi_\mathrm{in}}{n}\vert_{\partial S}=\pbyp{\psi_\mathrm{out}}{n}\vert_{\partial S}$ where $\mu = 1 / n^2$ for transverse electric (TE) polarization and $\mu = 1$ for transverse magnetic (TM) polarization. Then the wave function outside of the resonator is related to that on its boundary by
\begin{equation} \begin{array}{lcr} \psi_\mathrm{out}(\vec{r}) & = & \oint_{\partial S} \, \mathrm{d}|\vec{r}{\,'}| \bigg\{ \, \mu G(\vec{r}, \vec{r}{\,'}) \pbyp{\psi_\mathrm{in}}{n}(\vec{r}{\,'}) - \pbyp{G}{n}(\vec{r}, \vec{r}{\,'}) \psi_\mathrm{in}(\vec{r}{\,'}) \bigg\} \, . \end{array} \end{equation}

First, we choose as ansatz inside the resonator a plane wave $\psi_\mathrm{in}(\vec{r}) = \psi_0 \exp\{i (k_x x + k_y y) \}$ and calculate the corresponding wave function $\psi_\mathrm{out}(\vec r)$ at a location $\vec r=(r\cos\varphi , r\sin\varphi)$ far away from the domain $S$, i.e., for $k |\vec r-\vec r{\,'}| \gg 1$. Then the integral can be simplified by using $H_0^{(1)}(z= k |\vec{r} - \vec{r}{\,'}|) \approx \sqrt{2 / (\pi z)} \mathrm{e}^{i (z - \pi / 4)}$. Furthermore, in this limit $|\vec{r} - \vec{r}{\,'}| \simeq r - (x' \cos\varphi + y' \sin\varphi)$. The integration yields
\begin{equation}
\begin{array}{rcl}
 \psi_\mathrm{out}(r, \varphi) & = & -i \psi_0 \sqrt{\frac{2}{\pi k r}} \mathrm{e}^{i(k r - \pi / 4)} \sin\left[(k_x - k \cos{\varphi}) \frac{a}{2}\right] \\
 & & \sin\left[(k_y - k \sin{\varphi}) \frac{a}{2}\right] \left[ \frac{\mu k_x + k \cos{\varphi}}{k_y - k \sin{\varphi}} + \frac{\mu k_y + k \sin{\varphi}}{k_x - k \cos{\varphi}} \right]
\end{array} \end{equation}
where $a$ is the side length of the square.\\
It is instructive to express the wave vector components in terms of the angle of incidence, $(k_x,k_y)=n k(\cos{\alpha},\sin{\alpha})$ and to focus on the two terms in the last square bracket, which are maximal when their denominators vanish. The first term stems from the integration along the side walls parallel to the $y$ axis and is maximal for $n \sin{\alpha} = \sin{\varphi}$. This is simply Snell's law for a ray refracted at a side wall parallel to the $y$ axis. Analogously, the maximum of the second term corresponds to a ray refracted at a side wall parallel to the $x$ axis. 

The results for the full wave functions of the dielectric square are obtained by adding the $8$ plane waves with corresponding momentum vectors and relative amplitudes. For the symmetry class $A_2$ this yields
\begin{equation} \label{eq:ffModel} \psi_{A_2}(r, \varphi) = -i \frac{\psi_0}{2 \sqrt{2 \pi k r}} \mathrm{e}^{i(k r - \pi / 4)} g_{A_2}(\varphi) \end{equation}
with
\begin{equation} \scriptscriptstyle \label{eq:ffA2} \begin{array}{rcl} g_{A_2}(\varphi) & = & 2 \{ \sinc[(k_y - k \sin{\varphi}) \frac{a}{2}] \\
& & - \sinc[(-k_y - k \sin{\varphi}) \frac{a}{2}] \} \\
& & \times [\mu k_x \frac{a}{2} \cos(k_x \frac{a}{2}) \sin(k \frac{a}{2} \cos{\varphi}) \\ & & - k \frac{a}{2} \cos{\varphi} \cos(k \frac{a}{2} \cos{\varphi}) \sin(k_x \frac{a}{2})] \\ \\
& & +2 \{ \sinc[(k_x - k \cos{\varphi}) \frac{a}{2}] \\
& & - \sinc[(-k_x - k \cos{\varphi}) \frac{a}{2}] \} \\
& & \times [\mu k_y \frac{a}{2} \cos(k_y \frac{a}{2}) \sin(k \frac{a}{2} \sin{\varphi}) \\ & & - k \frac{a}{2} \sin{\varphi} \cos(k \frac{a}{2} \sin{\varphi}) \sin(k_y \frac{a}{2})] \\ \\
& & +2 \{ -\sinc[(k_x - k \sin{\varphi}) \frac{a}{2}] \\
& & + \sinc[(-k_x - k \sin{\varphi}) \frac{a}{2}] \} \\
& & \times [\mu k_y \frac{a}{2} \cos(k_y \frac{a}{2}) \sin(k \frac{a}{2} \cos{\varphi}) \\ & & - k \frac{a}{2} \cos{\varphi} \cos(k \frac{a}{2} \cos{\varphi}) \sin(k_y \frac{a}{2})] \\ \\
& & +2 \{ -\sinc[(k_y - k \cos{\varphi}) \frac{a}{2}] \\
& & + \sinc[(-k_y - k \cos{\varphi}) \frac{a}{2}] \} \\
& & \times [\mu k_x \frac{a}{2} \cos(k_x \frac{a}{2}) \sin(k \frac{a}{2} \sin{\varphi}) \\ & & - k \frac{a}{2} \sin{\varphi} \cos(k \frac{a}{2} \sin{\varphi}) \sin(k_x \frac{a}{2})] \, . \\ \\
\end{array} \end{equation}

\section{Effect of mode symmetries and degeneracies on the gain efficiency} \label{AppII}

Close to the laser threshold, we neglect mode competition and stimulated emission. Then the gain efficiency in a dye-doped polymer matrix is proportional to the product of the absorption probability of the pump photons and the emission probability of photons into a given lasing mode, averaged over the orientations $\Omega$ of the dye molecules \cite{Gozhyk2015}. Assuming that the emission and absorption depend solely on the dipole transition moments $\vec d_e$ and $\vec d_a$, these probabilities are proportional to  $|\vec{d}_e\cdot \vec{E}|^2$ and $|\vec{d}_a \cdot\vec{p} \, |^2$, respectively. Here, $\vec{E}$ denotes the electric field distribution and $\vec{p}=(p_x,p_y)$ is the unit vector describing the polarization state of the pump beam. Then, the probability $\mathcal{P}$ that the mode is excited is given by
\begin{equation} \mathcal{P}\propto\int_S \mathrm{d}S \int \mathrm{d}\Omega \,\rho(\Omega) \left| \vec{d}_e(\Omega)\cdot \vec{E} \right|^2 \left| \vec{d}_a(\Omega)\cdot\vec{p} \right|^2\end{equation}
where the integral is performed over the surface $S$ of the cavity, and $\rho(\Omega)$ is the distribution of orientations of the dye molecules. We assume that the lasing threshold of a mode is proportional to $1/\mathcal{P}$. In the case of TE polarization, the electric field is given by
\begin{equation} \vec{E} = -i \frac{\omega}{(n k)^2} \ez \times \vec\nabla B_z \propto \left( -\pbyp{\psi}{y}, \pbyp{\psi}{x}, 0 \right)^T \end{equation}
where the wave function $\psi$ corresponds to the $z$ component of the magnetic field, $B_z = B_0 \psi(x, y) \mathrm{e}^{-i \omega t}$ and $\omega = c k$ is the angular frequency of the mode. Furthermore, as outlined in the main text, we may assume that the angular distribution of the dye molecule orientations is uniform, $\rho(\Omega) = 1 / (4 \pi)$, and that $\vec{d}_e$ and $\vec{d}_a$ are parallel. This yields
\begin{equation} \label{eq:pumpEffUniform}
\begin{array}{c}
\mathcal{P} \propto \frac{(d_a d_e B_0 \omega)^2}{15 (n k)^4} \int_S \mathrm{d}S \bigg[ (2 |p_y|^2 + 1) \left| \pbyp{\psi}{x} \right|^2 \\ \\ + (2 |p_x|^2 + 1) \left| \pbyp{\psi}{y} \right|^2 - 4 \Re\left(\pbyp{\psi}{y} \pbyp{\psi^*}{x}\right) \Re\left(p_x p_y^*\right) \bigg] \, .
\end{array} \end{equation}
Note that under a linear pump polarization and for a plane wave, Eq.~(\ref{eq:pumpEffUniform}) yields a $3$ times higher pump efficiency if the wave propagates in the direction perpendicular to $\vec{p}$ compared to propagation parallel to it \cite{Gozhyk2015}. 

Equation (\ref{eq:pumpEffUniform}) in itself can only explain variations of the lasing threshold as a function of the pump polarization, but not changes of the emission characteristics and the symmetry of the lasing mode. This effect is attributed to the existence of pairs of modes that are degenerate but have different emission features. We assume that their superposition is determined by the maximum of the probability $\mathcal{P}$ for a given pump polarization. 

The square resonator exhibits $C_{4v}$ symmetry. Therefore, the resonance states belong to six different symmetry classes. These and the associated wave functions are summarized in Table \ref{tab:dlmWFs}. They form three pairs $\{S_1, S_2\}$ of degenerate modes, where we refer to $\{S_1, S_2\} = \{A_1, B_2\}$, $\{S_1, S_2\} = \{B_1, A_2\}$ and $\{S_1, S_2\} = \{E, E\}$  as the cases $\I,\, \II$ and $\III$, respectively. It should be noted that the two $E$ classes are degenerate due to symmetry reasons, whereas the degeneracy of the classes $A_1$ and $B_2$ and of the classes $B_1$ and $A_2$ is a particularity of the semiclassical model for the square resonator \cite{EPL-carre}, but not of the $C_{4v}$ point group in general. Numerical calculations show that resonances of these two pairs of symmetry classes are degenerate in very good approximation. 

In the following, we calculate the superpositions of degenerate mode pairs which maximize the probability $\mathcal{P}$, Eq.~(\ref{eq:pumpEffUniform}). A general superposition can be written as
\begin{equation} \psisup = \cos(\gamma)\, \psi_{S_1} + \mathrm{e}^{i \delta} \sin(\gamma)\, \psi_{S_2} \end{equation}
where the wave functions $\psi_{S_{1,2}}$ have equal amplitudes $\psi_0$. The whole space of possible superpositions is covered by the two parameters $\gamma \in [0, \pi)$ and $\delta \in [-\pi / 2, \pi / 2)$. In Ref.~\cite{EPL-carre}, we evidenced that the lasing modes have an angle of incidence $\alpha \geq \alpha_c$, where $\alpha_c$ denotes the critical angle. Therefore we may assume in good approximation that the wave vector components $k_{x, y}$ are real. 

We start with case $\I$. Since $\psi_{A_1}$ and $\psi_{B_2}$ are symmetric with respect to the $x$ and $y$ axes, also $\psisup$ has these symmetries. Hence the integral over the third term in the square brackets in~Eq.~(\ref{eq:pumpEffUniform}) vanishes. The remaining two terms yield after a lengthy but straightforward calculation
\begin{equation} \mathcal{P}_\mathrm{I} = \frac{(d_a d_e B_0 \omega)^2}{15 (n k)^2} |\psi_0|^2 \frac{a^2}{4} \mathcal{P}_\mathrm{\I, red} \end{equation}
where the dimensionless quantity $\mathcal{P}_\mathrm{\I, red}$ is called the reduced gain efficiency. It may be written as
\begin{equation} \label{eq:Pred3terms} \mathcal{P}_\mathrm{I, red} = \frac{1}{(n k)^2} [T_{\I, 1} + T_{\I, 2} + T_{\I, 3}] \end{equation}
where
\begin{equation} T_{\I, 1} = f_{\I, 1} [ 4 n^2 k^2 + 2 \cos(\delta) \sin(2 \gamma) (|p_y|^2 - |p_x|^2) (k_x^2 - k_y^2) ] \end{equation}
\begin{equation} T_{\I, 2} = f_{\I, 2} [ 4 (k_x^2 - k_y^2) + 2 n^2 k^2 \cos(\delta) \sin(2 \gamma) (|p_y|^2 - |p_x|^2) ] \end{equation}
\begin{equation} T_{\I, 3} = 8 f_{\I, 3} k_x k_y \cos(2 \gamma). \end{equation}
The terms $f_{\I, j}$ are given by
\begin{equation}f_{\I, 1} = 1 - \sinc(k_x a) \sinc(k_y a)\end{equation}
\begin{equation}f_{\I, 2} = \sinc(k_y a) - \sinc(k_x a)\end{equation}
\begin{equation}f_{\I, 3} = \sinc^2[(k_x - k_y) a/2] - \sinc^2[(k_x + k_y) a/2] \, . \end{equation}

The result for case $\II$ again may be written in the form Eq.~(\ref{eq:Pred3terms}) with $T_{\II, 1} = T_{\I, 1}$, $T_{\II, 2} = -T_{\I, 2}$ and $T_{\II, 3} = T_{\I, 3}$. For case $\III$, the third term in Eq.~(\ref{eq:pumpEffUniform}) does not vanish since $\psisup$ does not exhibit any symmetry. The result for the reduced gain efficiency has the same form as Eq.~(\ref{eq:Pred3terms}) but with four terms:
\begin{equation} T_{\III, 1} = f_{\III, 1} [ 4 n^2 k^2 + 2 \cos(\delta) \sin(2 \gamma) (|p_y|^2 - |p_x|^2) (k_x^2 - k_y^2) ] \end{equation}
\begin{equation} T_{\III, 2} = f_{\III, 2} [ 4 (k_x^2 - k_y^2) + 2 n^2 k^2 \cos(\delta) \sin(2 \gamma) (|p_y|^2 - |p_x|^2) ] \end{equation}
\begin{equation} T_{\III, 3} = -4 n^2 k^2 \cos(2 \gamma) \Re(p_x p_y^*) f_{\III, 3} \end{equation}
\begin{equation} T_{\III, 4} = -4 \cos(2 \gamma) \Re(p_x p_y^*) (k_x^2 - k_y^2) f_{\III, 4}. \end{equation}
The terms $f_{\III, j}$ are given by
\begin{equation} f_{\III, 1} = 1 + \sinc(k_x a) \sinc(k_y a)\end{equation}
\begin{equation} f_{\III, 2} = \sinc(k_x a) + \sinc(k_y a)\end{equation}
\begin{equation} f_{\III, 3} = \sinc^2[(k_x - k_y) a/2] + \sinc^2[(k_x + k_y) a/2]\end{equation}
\begin{equation} f_{\III, 4} = 2 \sinc[(k_x - k_y) a/2] \sinc[(k_x + k_y) a/2] \, . \end{equation}

These formulas can be further simplified by expressing the wave-vector components in terms of the angle of incidence, $(k_x,k_y)=n k(\cos\alpha, \sin\alpha)$. With the notation $\Delta p = |p_y|^2 - |p_x|^2$ this yields for case $\I$
\begin{equation} \begin{array}{rcl} \mathcal{P}_\mathrm{\I, red} & = & 2 \cos(\delta) \sin(2 \gamma) \Delta p \,[\cos(2 \alpha) f_\mathrm{I, 1} + f_{\I, 2}] \\
& & +4 \cos(2 \gamma) \sin(2 \alpha) f_\mathrm{\I, 3} \\
& & +4 [f_\mathrm{\I, 1} + \cos(2 \alpha) f_\mathrm{\I, 2}] \, . \end{array} \end{equation}
The result for $\mathcal{P}_\mathrm{\II, red}$ is obtained by replacing $f_\mathrm{\I, 2}$ by $-f_\mathrm{\I, 2}$ in $\mathcal{P}_\mathrm{\I, red}$. The result for case $\III$ is
\begin{equation} \begin{array}{rcl} \mathcal{P}_\mathrm{III, red} & = & f_\mathrm{\III, 1} [4 + 2 \cos(\delta) \sin(2 \gamma) \cos(2 \alpha) \Delta p] \\
 & & + f_\mathrm{\III, 2} [4 \cos(2 \alpha) + 2 \cos(\delta) \sin(2 \gamma) \Delta p] \\
 & & -4 \cos(2 \gamma) \Re(p_x p_y^*) [f_\mathrm{\III, 3} + f_\mathrm{\III, 4} \cos(2 \alpha)] \, . \end{array} \end{equation}

The next step is to calculate the extrema of $\mathcal{P}_\mathrm{red}$ as a function of the superposition parameters $\gamma$ and $\delta$. In all three cases, $\pbyp{\mathcal{P}_\mathrm{red}}{\delta} \propto \sin(\delta)$. Hence the only relevant value of $\delta$ is $\delta = 0$, which is implicitly assumed in the following calculations. The extremal points of $\mathcal{P}_\mathrm{red}$ can now be easily found by calculating the roots of $\pbyp{\mathcal{P}_\mathrm{I,red}}{\gamma}$. For case $\I$ the derivative with respect to $\gamma$ is
\begin{equation} \begin{array}{rcl} \pbyp{\mathcal{P}_\mathrm{I,red}}{\gamma} & = & 4 \cos(2 \gamma)  \Delta p [\cos(2 \alpha) f_{\I,1}(\alpha) + f_{\I,2}(\alpha)] \\ & & - 8 \sin(2 \gamma) \sin(2 \alpha) f_{\I,3}(\alpha) \, , \end{array} \end{equation}
and its roots are given by
\begin{equation} \label{eq:gammmaExtrIana} \cot(2 \gamma_\I) = \frac{1}{ \Delta p} \frac{2 \sin(2 \alpha) f_{I,3}}{\cos(2 \alpha) f_{\I,1} + f_{\I,2}} \, . \end{equation}
It should be noted that this is an exact formula as long as modes with $\alpha > \alpha_c$ are considered. Since the $\cot(z)$ is $\pi$-periodic, further roots can be found by adding $\pi/2$ to a root $\gamma_\I$. So in practice there is one maximum and one minimum, one in the interval $\gamma_\I \in [0, \pi/2]$ and the other one in the interval $\gamma_\I \in [\pi/2, \pi]$. For case $\II$ we obtain
\begin{equation} \cot(2 \gamma_\II) = \frac{1}{ \Delta p} \frac{2 \sin(2 \alpha) f_{I,3}}{\cos(2 \alpha) f_{\I,1} - f_{\I,2}} \, , \end{equation}
and for case $\III$
\begin{equation} \label{eq:caseIIIalphaRoot} \cot(2 \gamma_\III) = -2 \frac{\Re(p_x p_y^*)}{\Delta p} \frac{f_{\III,3}(\alpha) + \cos(2 \alpha) f_{\III,4}(\alpha)}{f_{\III,1}(\alpha) \cos(2 \alpha) + f_{\III,2}(\alpha)} \, . \end{equation}

The quantities $f$ feature several terms of the form $\textrm{sinc}(n k_{x, y} a)$. For a square with side length $a = 120~\mu$m and emission wavelength $\lambda$ around $600$~nm, $n k a \simeq 1900$. So all these terms are of the order of $10^{-3}$ or smaller, and can be neglected. Thus, in the semiclassical regime, the expressions for $\mathcal{P}$ can be further simplified. 


\begin{thebibliography}{39}%
\makeatletter
\providecommand \@ifxundefined [1]{%
 \@ifx{#1\undefined}
}%
\providecommand \@ifnum [1]{%
 \ifnum #1\expandafter \@firstoftwo
 \else \expandafter \@secondoftwo
 \fi
}%
\providecommand \@ifx [1]{%
 \ifx #1\expandafter \@firstoftwo
 \else \expandafter \@secondoftwo
 \fi
}%
\providecommand \natexlab [1]{#1}%
\providecommand \enquote  [1]{``#1''}%
\providecommand \bibnamefont  [1]{#1}%
\providecommand \bibfnamefont [1]{#1}%
\providecommand \citenamefont [1]{#1}%
\providecommand \href@noop [0]{\@secondoftwo}%
\providecommand \href [0]{\begingroup \@sanitize@url \@href}%
\providecommand \@href[1]{\@@startlink{#1}\@@href}%
\providecommand \@@href[1]{\endgroup#1\@@endlink}%
\providecommand \@sanitize@url [0]{\catcode `\\12\catcode `\$12\catcode
  `\&12\catcode `\#12\catcode `\^12\catcode `\_12\catcode `\%12\relax}%
\providecommand \@@startlink[1]{}%
\providecommand \@@endlink[0]{}%
\providecommand \url  [0]{\begingroup\@sanitize@url \@url }%
\providecommand \@url [1]{\endgroup\@href {#1}{\urlprefix }}%
\providecommand \urlprefix  [0]{URL }%
\providecommand \Eprint [0]{\href }%
\providecommand \doibase [0]{http://dx.doi.org/}%
\providecommand \selectlanguage [0]{\@gobble}%
\providecommand \bibinfo  [0]{\@secondoftwo}%
\providecommand \bibfield  [0]{\@secondoftwo}%
\providecommand \translation [1]{[#1]}%
\providecommand \BibitemOpen [0]{}%
\providecommand \bibitemStop [0]{}%
\providecommand \bibitemNoStop [0]{.\EOS\space}%
\providecommand \EOS [0]{\spacefactor3000\relax}%
\providecommand \BibitemShut  [1]{\csname bibitem#1\endcsname}%
\let\auto@bib@innerbib\@empty
\bibitem [{\citenamefont {Matsko}(2009)}]{matsko-livre}%
  \BibitemOpen
  \bibinfo {editor} {\bibfnamefont {A.~B.}\ \bibnamefont {Matsko}},\ ed.,\
  \href@noop {} {\emph {\bibinfo {title} {Practical Applications of
  Microresonators in Optics and Photonics}}}\ (\bibinfo  {publisher} {CRC
  Press},\ \bibinfo {address} {Boca Raton, FL},\ \bibinfo {year}
  {2009})\BibitemShut {NoStop}%
\bibitem [{\citenamefont {Foreman}\ \emph {et~al.}(2015)\citenamefont
  {Foreman}, \citenamefont {Swaim},\ and\ \citenamefont
  {Vollmer}}]{revue-capteur-bio}%
  \BibitemOpen
  \bibfield  {author} {\bibinfo {author} {\bibfnamefont {M.~R.}\ \bibnamefont
  {Foreman}}, \bibinfo {author} {\bibfnamefont {J.~D.}\ \bibnamefont {Swaim}},
  \ and\ \bibinfo {author} {\bibfnamefont {F.}~\bibnamefont {Vollmer}},\ }\href
  {\doibase 10.1364/AOP.7.000168} {\bibfield  {journal} {\bibinfo  {journal}
  {Adv. Opt. Photon.}\ }\textbf {\bibinfo {volume} {7}},\ \bibinfo {pages}
  {168} (\bibinfo {year} {2015})}\BibitemShut {NoStop}%
\bibitem [{\citenamefont {Wade}\ and\ \citenamefont
  {Bailey}(2016)}]{revue-capteur-chimie}%
  \BibitemOpen
  \bibfield  {author} {\bibinfo {author} {\bibfnamefont {J.~H.}\ \bibnamefont
  {Wade}}\ and\ \bibinfo {author} {\bibfnamefont {R.~C.}\ \bibnamefont
  {Bailey}},\ }\href {\doibase 10.1146/annurev-anchem-071015-041742} {\bibfield
   {journal} {\bibinfo  {journal} {Annu. Rev. Anal. Chem.}\ }\textbf {\bibinfo
  {volume} {9}},\ \bibinfo {pages} {1} (\bibinfo {year} {2016})}\BibitemShut
  {NoStop}%
\bibitem [{\citenamefont {Schwefel}\ \emph {et~al.}(2004)\citenamefont
  {Schwefel}, \citenamefont {Rex}, \citenamefont {Tureci}, \citenamefont
  {Chang}, \citenamefont {Stone}, \citenamefont {Ben-Messaoud},\ and\
  \citenamefont {Zyss}}]{schwefel}%
  \BibitemOpen
  \bibfield  {author} {\bibinfo {author} {\bibfnamefont {H.~G.~L.}\
  \bibnamefont {Schwefel}}, \bibinfo {author} {\bibfnamefont {N.~B.}\
  \bibnamefont {Rex}}, \bibinfo {author} {\bibfnamefont {H.~E.}\ \bibnamefont
  {Tureci}}, \bibinfo {author} {\bibfnamefont {R.~K.}\ \bibnamefont {Chang}},
  \bibinfo {author} {\bibfnamefont {A.~D.}\ \bibnamefont {Stone}}, \bibinfo
  {author} {\bibfnamefont {T.}~\bibnamefont {Ben-Messaoud}}, \ and\ \bibinfo
  {author} {\bibfnamefont {J.}~\bibnamefont {Zyss}},\ }\href {\doibase
  10.1364/JOSAB.21.000923} {\bibfield  {journal} {\bibinfo  {journal} {J. Opt.
  Soc. Am. B}\ }\textbf {\bibinfo {volume} {21}},\ \bibinfo {pages} {923}
  (\bibinfo {year} {2004})}\BibitemShut {NoStop}%
\bibitem [{\citenamefont {Shinohara}\ \emph {et~al.}(2009)\citenamefont
  {Shinohara}, \citenamefont {Hentschel}, \citenamefont {Wiersig},
  \citenamefont {Sasaki},\ and\ \citenamefont {Harayama}}]{limacon}%
  \BibitemOpen
  \bibfield  {author} {\bibinfo {author} {\bibfnamefont {S.}~\bibnamefont
  {Shinohara}}, \bibinfo {author} {\bibfnamefont {M.}~\bibnamefont
  {Hentschel}}, \bibinfo {author} {\bibfnamefont {J.}~\bibnamefont {Wiersig}},
  \bibinfo {author} {\bibfnamefont {T.}~\bibnamefont {Sasaki}}, \ and\ \bibinfo
  {author} {\bibfnamefont {T.}~\bibnamefont {Harayama}},\ }\href {\doibase
  10.1103/PhysRevA.80.031801} {\bibfield  {journal} {\bibinfo  {journal} {Phys.
  Rev. A}\ }\textbf {\bibinfo {volume} {80}},\ \bibinfo {pages} {031801}
  (\bibinfo {year} {2009})}\BibitemShut {NoStop}%
\bibitem [{\citenamefont {Lebental}\ \emph
  {et~al.}(2007{\natexlab{a}})\citenamefont {Lebental}, \citenamefont {Lauret},
  \citenamefont {Zyss}, \citenamefont {Schmit},\ and\ \citenamefont
  {Bogomolny}}]{lebental-stade-PRA}%
  \BibitemOpen
  \bibfield  {author} {\bibinfo {author} {\bibfnamefont {M.}~\bibnamefont
  {Lebental}}, \bibinfo {author} {\bibfnamefont {J.~S.}\ \bibnamefont
  {Lauret}}, \bibinfo {author} {\bibfnamefont {J.}~\bibnamefont {Zyss}},
  \bibinfo {author} {\bibfnamefont {C.}~\bibnamefont {Schmit}}, \ and\ \bibinfo
  {author} {\bibfnamefont {E.}~\bibnamefont {Bogomolny}},\ }\href {\doibase
  10.1103/PhysRevA.75.033806} {\bibfield  {journal} {\bibinfo  {journal} {Phys.
  Rev. A}\ }\textbf {\bibinfo {volume} {75}},\ \bibinfo {pages} {033806}
  (\bibinfo {year} {2007}{\natexlab{a}})}\BibitemShut {NoStop}%
\bibitem [{\citenamefont {Knitter}\ \emph {et~al.}(2013)\citenamefont
  {Knitter}, \citenamefont {Kues}, \citenamefont {Haidl},\ and\ \citenamefont
  {Fallnich}}]{pola-aleatoire}%
  \BibitemOpen
  \bibfield  {author} {\bibinfo {author} {\bibfnamefont {S.}~\bibnamefont
  {Knitter}}, \bibinfo {author} {\bibfnamefont {M.}~\bibnamefont {Kues}},
  \bibinfo {author} {\bibfnamefont {M.}~\bibnamefont {Haidl}}, \ and\ \bibinfo
  {author} {\bibfnamefont {C.}~\bibnamefont {Fallnich}},\ }\href {\doibase
  10.1364/OE.21.031591} {\bibfield  {journal} {\bibinfo  {journal} {Opt.
  Express}\ }\textbf {\bibinfo {volume} {21}},\ \bibinfo {pages} {31591}
  (\bibinfo {year} {2013})}\BibitemShut {NoStop}%
\bibitem [{\citenamefont {Yu}\ \emph {et~al.}(2017)\citenamefont {Yu},
  \citenamefont {Ding}, \citenamefont {Zhang},\ and\ \citenamefont
  {Zhang}}]{JAP-theorie-QD}%
  \BibitemOpen
  \bibfield  {author} {\bibinfo {author} {\bibfnamefont {R.}~\bibnamefont
  {Yu}}, \bibinfo {author} {\bibfnamefont {C.}~\bibnamefont {Ding}}, \bibinfo
  {author} {\bibfnamefont {D.}~\bibnamefont {Zhang}}, \ and\ \bibinfo {author}
  {\bibfnamefont {S.}~\bibnamefont {Zhang}},\ }\href {\doibase
  10.1063/1.4981130} {\bibfield  {journal} {\bibinfo  {journal} {J. Appl.
  Phys}\ }\textbf {\bibinfo {volume} {121}},\ \bibinfo {pages} {144303}
  (\bibinfo {year} {2017})}\BibitemShut {NoStop}%
\bibitem [{\citenamefont {Hu}(2017)}]{srep-spintransistor}%
  \BibitemOpen
  \bibfield  {author} {\bibinfo {author} {\bibfnamefont {C.~Y.}\ \bibnamefont
  {Hu}},\ }\href {\doibase 10.1038/srep45582} {\bibfield  {journal} {\bibinfo
  {journal} {Scientific Reports}\ }\textbf {\bibinfo {volume} {7}},\ \bibinfo
  {pages} {45582} (\bibinfo {year} {2017})}\BibitemShut {NoStop}%
\bibitem [{\citenamefont {Ben-Messaoud}\ and\ \citenamefont
  {Zyss}(2005)}]{tahar}%
  \BibitemOpen
  \bibfield  {author} {\bibinfo {author} {\bibfnamefont {T.}~\bibnamefont
  {Ben-Messaoud}}\ and\ \bibinfo {author} {\bibfnamefont {J.}~\bibnamefont
  {Zyss}},\ }\href {\doibase DOI:10.1063/1.1949708} {\bibfield  {journal}
  {\bibinfo  {journal} {Appl. Phys. Lett.}\ }\textbf {\bibinfo {volume} {86}},\
  \bibinfo {pages} {241110} (\bibinfo {year} {2005})}\BibitemShut {NoStop}%
\bibitem [{\citenamefont {Bachelard}\ \emph {et~al.}(2014)\citenamefont
  {Bachelard}, \citenamefont {Gigan}, \citenamefont {Noblin},\ and\
  \citenamefont {Sebbah}}]{sebbah}%
  \BibitemOpen
  \bibfield  {author} {\bibinfo {author} {\bibfnamefont {N.}~\bibnamefont
  {Bachelard}}, \bibinfo {author} {\bibfnamefont {S.}~\bibnamefont {Gigan}},
  \bibinfo {author} {\bibfnamefont {X.}~\bibnamefont {Noblin}}, \ and\ \bibinfo
  {author} {\bibfnamefont {P.}~\bibnamefont {Sebbah}},\ }\href {\doibase
  10.1038/nphys2939} {\bibfield  {journal} {\bibinfo  {journal} {Nature
  Physics}\ }\textbf {\bibinfo {volume} {10}},\ \bibinfo {pages} {426}
  (\bibinfo {year} {2014})}\BibitemShut {NoStop}%
\bibitem [{\citenamefont {Liew}\ \emph {et~al.}(2014)\citenamefont {Liew},
  \citenamefont {Redding}, \citenamefont {Ge}, \citenamefont {Solomon},\ and\
  \citenamefont {Cao}}]{huicao-disk}%
  \BibitemOpen
  \bibfield  {author} {\bibinfo {author} {\bibfnamefont {S.~F.}\ \bibnamefont
  {Liew}}, \bibinfo {author} {\bibfnamefont {B.}~\bibnamefont {Redding}},
  \bibinfo {author} {\bibfnamefont {L.}~\bibnamefont {Ge}}, \bibinfo {author}
  {\bibfnamefont {G.~S.}\ \bibnamefont {Solomon}}, \ and\ \bibinfo {author}
  {\bibfnamefont {H.}~\bibnamefont {Cao}},\ }\href {\doibase 10.1063/1.4883637}
  {\bibfield  {journal} {\bibinfo  {journal} {Appl. Phys. Lett.}\ }\textbf
  {\bibinfo {volume} {104}},\ \bibinfo {pages} {231108} (\bibinfo {year}
  {2014})}\BibitemShut {NoStop}%
\bibitem [{\citenamefont {Thyrrestrup}\ \emph {et~al.}(2014)\citenamefont
  {Thyrrestrup}, \citenamefont {Y\"uce}, \citenamefont {Ctistis}, \citenamefont
  {Claudon}, \citenamefont {Vos},\ and\ \citenamefont {G\'erard}}]{JMGerard}%
  \BibitemOpen
  \bibfield  {author} {\bibinfo {author} {\bibfnamefont {H.}~\bibnamefont
  {Thyrrestrup}}, \bibinfo {author} {\bibfnamefont {E.}~\bibnamefont {Y\"uce}},
  \bibinfo {author} {\bibfnamefont {G.}~\bibnamefont {Ctistis}}, \bibinfo
  {author} {\bibfnamefont {J.}~\bibnamefont {Claudon}}, \bibinfo {author}
  {\bibfnamefont {W.~L.}\ \bibnamefont {Vos}}, \ and\ \bibinfo {author}
  {\bibfnamefont {J.-M.}\ \bibnamefont {G\'erard}},\ }\href {\doibase
  10.1063/1.4896160} {\bibfield  {journal} {\bibinfo  {journal} {Appl. Phys.
  Lett.}\ }\textbf {\bibinfo {volume} {105}},\ \bibinfo {pages} {111115}
  (\bibinfo {year} {2014})}\BibitemShut {NoStop}%
\bibitem [{\citenamefont {Estrecho}\ \emph {et~al.}(2016)\citenamefont
  {Estrecho}, \citenamefont {Gao}, \citenamefont {Brodbeck}, \citenamefont
  {Kamp}, \citenamefont {Schneider}, \citenamefont {H\"ofling}, \citenamefont
  {Truscott},\ and\ \citenamefont {Ostrovskaya}}]{srep-polariton-Berry}%
  \BibitemOpen
  \bibfield  {author} {\bibinfo {author} {\bibfnamefont {E.}~\bibnamefont
  {Estrecho}}, \bibinfo {author} {\bibfnamefont {T.}~\bibnamefont {Gao}},
  \bibinfo {author} {\bibfnamefont {S.}~\bibnamefont {Brodbeck}}, \bibinfo
  {author} {\bibfnamefont {M.}~\bibnamefont {Kamp}}, \bibinfo {author}
  {\bibfnamefont {C.}~\bibnamefont {Schneider}}, \bibinfo {author}
  {\bibfnamefont {S.}~\bibnamefont {H\"ofling}}, \bibinfo {author}
  {\bibfnamefont {A.~G.}\ \bibnamefont {Truscott}}, \ and\ \bibinfo {author}
  {\bibfnamefont {E.~A.}\ \bibnamefont {Ostrovskaya}},\ }\href {\doibase
  10.1038/srep37653} {\bibfield  {journal} {\bibinfo  {journal} {Scientific
  Reports}\ }\textbf {\bibinfo {volume} {6}},\ \bibinfo {pages} {37653}
  (\bibinfo {year} {2016})}\BibitemShut {NoStop}%
\bibitem [{\citenamefont {Czerniuk}\ \emph {et~al.}(2017)\citenamefont
  {Czerniuk}, \citenamefont {Wigger}, \citenamefont {Akimov}, \citenamefont
  {Schneider}, \citenamefont {Kamp}, \citenamefont {H\"ofling}, \citenamefont
  {Yakovlev}, \citenamefont {Kuhn}, \citenamefont {Reiter},\ and\ \citenamefont
  {Bayer}}]{PRL-phonons}%
  \BibitemOpen
  \bibfield  {author} {\bibinfo {author} {\bibfnamefont {T.}~\bibnamefont
  {Czerniuk}}, \bibinfo {author} {\bibfnamefont {D.}~\bibnamefont {Wigger}},
  \bibinfo {author} {\bibfnamefont {A.~V.}\ \bibnamefont {Akimov}}, \bibinfo
  {author} {\bibfnamefont {C.}~\bibnamefont {Schneider}}, \bibinfo {author}
  {\bibfnamefont {M.}~\bibnamefont {Kamp}}, \bibinfo {author} {\bibfnamefont
  {S.}~\bibnamefont {H\"ofling}}, \bibinfo {author} {\bibfnamefont {D.~R.}\
  \bibnamefont {Yakovlev}}, \bibinfo {author} {\bibfnamefont {T.}~\bibnamefont
  {Kuhn}}, \bibinfo {author} {\bibfnamefont {D.~E.}\ \bibnamefont {Reiter}}, \
  and\ \bibinfo {author} {\bibfnamefont {M.}~\bibnamefont {Bayer}},\ }\href
  {\doibase 10.1103/PhysRevLett.118.133901} {\bibfield  {journal} {\bibinfo
  {journal} {Phys. Rev. Lett.}\ }\textbf {\bibinfo {volume} {118}},\ \bibinfo
  {pages} {133901} (\bibinfo {year} {2017})}\BibitemShut {NoStop}%
\bibitem [{\citenamefont {Lafont}\ \emph {et~al.}(2017)\citenamefont {Lafont},
  \citenamefont {Luk}, \citenamefont {Lewandowski}, \citenamefont {Kwong},
  \citenamefont {Leung}, \citenamefont {Galopin}, \citenamefont {Lemaitre},
  \citenamefont {Tignon}, \citenamefont {Schumacher}, \citenamefont {Baudin},\
  and\ \citenamefont {Binder}}]{APL-hall-optique}%
  \BibitemOpen
  \bibfield  {author} {\bibinfo {author} {\bibfnamefont {O.}~\bibnamefont
  {Lafont}}, \bibinfo {author} {\bibfnamefont {S.~M.~H.}\ \bibnamefont {Luk}},
  \bibinfo {author} {\bibfnamefont {P.}~\bibnamefont {Lewandowski}}, \bibinfo
  {author} {\bibfnamefont {N.~H.}\ \bibnamefont {Kwong}}, \bibinfo {author}
  {\bibfnamefont {P.~T.}\ \bibnamefont {Leung}}, \bibinfo {author}
  {\bibfnamefont {E.}~\bibnamefont {Galopin}}, \bibinfo {author} {\bibfnamefont
  {A.}~\bibnamefont {Lemaitre}}, \bibinfo {author} {\bibfnamefont
  {J.}~\bibnamefont {Tignon}}, \bibinfo {author} {\bibfnamefont
  {S.}~\bibnamefont {Schumacher}}, \bibinfo {author} {\bibfnamefont
  {E.}~\bibnamefont {Baudin}}, \ and\ \bibinfo {author} {\bibfnamefont
  {R.}~\bibnamefont {Binder}},\ }\href {\doibase 10.1063/1.4975681} {\bibfield
  {journal} {\bibinfo  {journal} {Appl. Phys. Lett.}\ }\textbf {\bibinfo
  {volume} {110}},\ \bibinfo {pages} {061108} (\bibinfo {year}
  {2017})}\BibitemShut {NoStop}%
\bibitem [{\citenamefont {Bittner}\ \emph {et~al.}(2016)\citenamefont
  {Bittner}, \citenamefont {Lafargue}, \citenamefont {Gozhyk}, \citenamefont
  {Djellali}, \citenamefont {Milliet}, \citenamefont {Hickox-Young},
  \citenamefont {Ulysse}, \citenamefont {Bouche}, \citenamefont {Dubertrand},
  \citenamefont {Bogomolny}, \citenamefont {Zyss},\ and\ \citenamefont
  {Lebental}}]{EPL-carre}%
  \BibitemOpen
  \bibfield  {author} {\bibinfo {author} {\bibfnamefont {S.}~\bibnamefont
  {Bittner}}, \bibinfo {author} {\bibfnamefont {C.}~\bibnamefont {Lafargue}},
  \bibinfo {author} {\bibfnamefont {I.}~\bibnamefont {Gozhyk}}, \bibinfo
  {author} {\bibfnamefont {N.}~\bibnamefont {Djellali}}, \bibinfo {author}
  {\bibfnamefont {L.}~\bibnamefont {Milliet}}, \bibinfo {author} {\bibfnamefont
  {D.~T.}\ \bibnamefont {Hickox-Young}}, \bibinfo {author} {\bibfnamefont
  {C.}~\bibnamefont {Ulysse}}, \bibinfo {author} {\bibfnamefont
  {D.}~\bibnamefont {Bouche}}, \bibinfo {author} {\bibfnamefont
  {R.}~\bibnamefont {Dubertrand}}, \bibinfo {author} {\bibfnamefont
  {E.}~\bibnamefont {Bogomolny}}, \bibinfo {author} {\bibfnamefont
  {J.}~\bibnamefont {Zyss}}, \ and\ \bibinfo {author} {\bibfnamefont
  {M.}~\bibnamefont {Lebental}},\ }\href {\doibase 10.1209/0295-5075/113/54002}
  {\bibfield  {journal} {\bibinfo  {journal} {Europhys. Lett.}\ }\textbf
  {\bibinfo {volume} {113}},\ \bibinfo {pages} {54002} (\bibinfo {year}
  {2016})}\BibitemShut {NoStop}%
\bibitem [{\citenamefont {Yang}\ and\ \citenamefont
  {Huang}(2016)}]{huang-carre}%
  \BibitemOpen
  \bibfield  {author} {\bibinfo {author} {\bibfnamefont {Y.-D.}\ \bibnamefont
  {Yang}}\ and\ \bibinfo {author} {\bibfnamefont {Y.-Z.}\ \bibnamefont
  {Huang}},\ }\href {\doibase 10.1088/0022-3727/49/25/253001} {\bibfield
  {journal} {\bibinfo  {journal} {J. Phys. D}\ }\textbf {\bibinfo {volume}
  {49}},\ \bibinfo {pages} {253001} (\bibinfo {year} {2016})}\BibitemShut
  {NoStop}%
\bibitem [{\citenamefont {Bittner}\ \emph {et~al.}(2013)\citenamefont
  {Bittner}, \citenamefont {Bogomolny}, \citenamefont {Dietz}, \citenamefont
  {Miski-Oglu},\ and\ \citenamefont {Richter}}]{Bittner2013}%
  \BibitemOpen
  \bibfield  {author} {\bibinfo {author} {\bibfnamefont {S.}~\bibnamefont
  {Bittner}}, \bibinfo {author} {\bibfnamefont {E.}~\bibnamefont {Bogomolny}},
  \bibinfo {author} {\bibfnamefont {B.}~\bibnamefont {Dietz}}, \bibinfo
  {author} {\bibfnamefont {M.}~\bibnamefont {Miski-Oglu}}, \ and\ \bibinfo
  {author} {\bibfnamefont {A.}~\bibnamefont {Richter}},\ }\href {\doibase
  10.1103/PhysRevE.88.062906} {\bibfield  {journal} {\bibinfo  {journal} {Phys.
  Rev. E}\ }\textbf {\bibinfo {volume} {88}},\ \bibinfo {pages} {062906}
  (\bibinfo {year} {2013})}\BibitemShut {NoStop}%
\bibitem [{\citenamefont {Tinkham}(1992)}]{tinkham}%
  \BibitemOpen
  \bibfield  {author} {\bibinfo {author} {\bibfnamefont {M.}~\bibnamefont
  {Tinkham}},\ }\href@noop {} {\emph {\bibinfo {title} {Group Theory and
  Quantum Mechanics}}}\ (\bibinfo  {publisher} {McGraw-Hill},\ \bibinfo {year}
  {1992})\BibitemShut {NoStop}%
\bibitem [{\citenamefont {Chen}\ \emph {et~al.}(2014)\citenamefont {Chen},
  \citenamefont {Sobeshchuk}, \citenamefont {Lafargue}, \citenamefont
  {Mansfield}, \citenamefont {Yom}, \citenamefont {Johnstone}, \citenamefont
  {Hales}, \citenamefont {Bittner}, \citenamefont {Charpignon}, \citenamefont
  {Ulbricht}, \citenamefont {Lautru}, \citenamefont {Denisyuk}, \citenamefont
  {Zyss}, \citenamefont {Perry},\ and\ \citenamefont {Lebental}}]{OE-3D}%
  \BibitemOpen
  \bibfield  {author} {\bibinfo {author} {\bibfnamefont {V.~W.}\ \bibnamefont
  {Chen}}, \bibinfo {author} {\bibfnamefont {N.}~\bibnamefont {Sobeshchuk}},
  \bibinfo {author} {\bibfnamefont {C.}~\bibnamefont {Lafargue}}, \bibinfo
  {author} {\bibfnamefont {E.~S.}\ \bibnamefont {Mansfield}}, \bibinfo {author}
  {\bibfnamefont {J.}~\bibnamefont {Yom}}, \bibinfo {author} {\bibfnamefont
  {L.}~\bibnamefont {Johnstone}}, \bibinfo {author} {\bibfnamefont {J.~M.}\
  \bibnamefont {Hales}}, \bibinfo {author} {\bibfnamefont {S.}~\bibnamefont
  {Bittner}}, \bibinfo {author} {\bibfnamefont {S.}~\bibnamefont {Charpignon}},
  \bibinfo {author} {\bibfnamefont {D.}~\bibnamefont {Ulbricht}}, \bibinfo
  {author} {\bibfnamefont {J.}~\bibnamefont {Lautru}}, \bibinfo {author}
  {\bibfnamefont {I.}~\bibnamefont {Denisyuk}}, \bibinfo {author}
  {\bibfnamefont {J.}~\bibnamefont {Zyss}}, \bibinfo {author} {\bibfnamefont
  {J.~W.}\ \bibnamefont {Perry}}, \ and\ \bibinfo {author} {\bibfnamefont
  {M.}~\bibnamefont {Lebental}},\ }\href {\doibase 10.1364/OE.22.012316}
  {\bibfield  {journal} {\bibinfo  {journal} {Opt. Express}\ }\textbf {\bibinfo
  {volume} {22}},\ \bibinfo {pages} {12316} (\bibinfo {year}
  {2014})}\BibitemShut {NoStop}%
\bibitem [{\citenamefont {Lafargue}\ \emph {et~al.}(2013)\citenamefont
  {Lafargue}, \citenamefont {Bittner}, \citenamefont {Lozenko}, \citenamefont
  {Lautru}, \citenamefont {Zyss}, \citenamefont {Ulysse}, \citenamefont
  {Cluzel},\ and\ \citenamefont {Lebental}}]{APL-FP}%
  \BibitemOpen
  \bibfield  {author} {\bibinfo {author} {\bibfnamefont {C.}~\bibnamefont
  {Lafargue}}, \bibinfo {author} {\bibfnamefont {S.}~\bibnamefont {Bittner}},
  \bibinfo {author} {\bibfnamefont {S.}~\bibnamefont {Lozenko}}, \bibinfo
  {author} {\bibfnamefont {J.}~\bibnamefont {Lautru}}, \bibinfo {author}
  {\bibfnamefont {J.}~\bibnamefont {Zyss}}, \bibinfo {author} {\bibfnamefont
  {C.}~\bibnamefont {Ulysse}}, \bibinfo {author} {\bibfnamefont
  {C.}~\bibnamefont {Cluzel}}, \ and\ \bibinfo {author} {\bibfnamefont
  {M.}~\bibnamefont {Lebental}},\ }\href {\doibase 10.1063/1.4812667}
  {\bibfield  {journal} {\bibinfo  {journal} {Appl. Phys. Lett.}\ }\textbf
  {\bibinfo {volume} {102}},\ \bibinfo {pages} {251120} (\bibinfo {year}
  {2013})}\BibitemShut {NoStop}%
\bibitem [{\citenamefont {Lebental}\ \emph
  {et~al.}(2007{\natexlab{b}})\citenamefont {Lebental}, \citenamefont
  {Djellali}, \citenamefont {Arnaud}, \citenamefont {Lauret}, \citenamefont
  {Zyss}, \citenamefont {Dubertrand}, \citenamefont {Schmit},\ and\
  \citenamefont {Bogomolny}}]{PRA-spectres}%
  \BibitemOpen
  \bibfield  {author} {\bibinfo {author} {\bibfnamefont {M.}~\bibnamefont
  {Lebental}}, \bibinfo {author} {\bibfnamefont {N.}~\bibnamefont {Djellali}},
  \bibinfo {author} {\bibfnamefont {C.}~\bibnamefont {Arnaud}}, \bibinfo
  {author} {\bibfnamefont {J.-S.}\ \bibnamefont {Lauret}}, \bibinfo {author}
  {\bibfnamefont {J.}~\bibnamefont {Zyss}}, \bibinfo {author} {\bibfnamefont
  {R.}~\bibnamefont {Dubertrand}}, \bibinfo {author} {\bibfnamefont
  {C.}~\bibnamefont {Schmit}}, \ and\ \bibinfo {author} {\bibfnamefont
  {E.}~\bibnamefont {Bogomolny}},\ }\href {\doibase 10.1103/PhysRevA.76.023830}
  {\bibfield  {journal} {\bibinfo  {journal} {Phys. Rev. A}\ }\textbf {\bibinfo
  {volume} {76}},\ \bibinfo {pages} {023830} (\bibinfo {year}
  {2007}{\natexlab{b}})}\BibitemShut {NoStop}%
\bibitem [{\citenamefont {Gennarelli}\ \emph {et~al.}(2015)\citenamefont
  {Gennarelli}, \citenamefont {Frongillo},\ and\ \citenamefont
  {Riccio}}]{gennarelli-2015}%
  \BibitemOpen
  \bibfield  {author} {\bibinfo {author} {\bibfnamefont {G.}~\bibnamefont
  {Gennarelli}}, \bibinfo {author} {\bibfnamefont {M.}~\bibnamefont
  {Frongillo}}, \ and\ \bibinfo {author} {\bibfnamefont {G.}~\bibnamefont
  {Riccio}},\ }\href {\doibase 10.1109/TAP.2014.2364305} {\bibfield  {journal}
  {\bibinfo  {journal} {IEEE Trans. Antennas Propag.}\ }\textbf {\bibinfo
  {volume} {63}},\ \bibinfo {pages} {374} (\bibinfo {year} {2015})}\BibitemShut
  {NoStop}%
\bibitem [{\citenamefont {McIsaac}(1975)}]{McIsaac}%
  \BibitemOpen
  \bibfield  {author} {\bibinfo {author} {\bibfnamefont {P.}~\bibnamefont
  {McIsaac}},\ }\href {\doibase 10.1109/TMTT.1975.1128584} {\bibfield
  {journal} {\bibinfo  {journal} {IEEE Trans. Microwave Theory Tech.}\ }\textbf
  {\bibinfo {volume} {23}},\ \bibinfo {pages} {421} (\bibinfo {year}
  {1975})}\BibitemShut {NoStop}%
\bibitem [{\citenamefont {Valeur}(2001)}]{valeur}%
  \BibitemOpen
  \bibfield  {author} {\bibinfo {author} {\bibfnamefont {B.}~\bibnamefont
  {Valeur}},\ }\href@noop {} {\emph {\bibinfo {title} {Molecular fluorescence:
  Principles and Applications}}}\ (\bibinfo  {publisher} {Wiley-VCH},\ \bibinfo
  {year} {2001})\BibitemShut {NoStop}%
\bibitem [{\citenamefont {Lakowicz}(2006)}]{lakowicz}%
  \BibitemOpen
  \bibfield  {author} {\bibinfo {author} {\bibfnamefont {J.~R.}\ \bibnamefont
  {Lakowicz}},\ }\href@noop {} {\emph {\bibinfo {title} {Principles of
  Fluorescence Spectroscopy}}},\ \bibinfo {edition} {3rd}\ ed.\ (\bibinfo
  {publisher} {Springer},\ \bibinfo {year} {2006})\BibitemShut {NoStop}%
\bibitem [{\citenamefont {Gozhyk}\ \emph {et~al.}(2012)\citenamefont {Gozhyk},
  \citenamefont {Clavier}, \citenamefont {M\'eallet-Renault}, \citenamefont
  {Dvor\-ko}, \citenamefont {Pansu}, \citenamefont {Audibert}, \citenamefont
  {Brosseau}, \citenamefont {Lafargue}, \citenamefont {Tsvirkun}, \citenamefont
  {Lozenko}, \citenamefont {Forget}, \citenamefont {Ch\'enais}, \citenamefont
  {Ulysse}, \citenamefont {Zyss},\ and\ \citenamefont
  {Lebental}}]{iryna-polarisation}%
  \BibitemOpen
  \bibfield  {author} {\bibinfo {author} {\bibfnamefont {I.}~\bibnamefont
  {Gozhyk}}, \bibinfo {author} {\bibfnamefont {G.}~\bibnamefont {Clavier}},
  \bibinfo {author} {\bibfnamefont {R.}~\bibnamefont {M\'eallet-Renault}},
  \bibinfo {author} {\bibfnamefont {M.}~\bibnamefont {Dvor\-ko}}, \bibinfo
  {author} {\bibfnamefont {R.}~\bibnamefont {Pansu}}, \bibinfo {author}
  {\bibfnamefont {J.-F.}\ \bibnamefont {Audibert}}, \bibinfo {author}
  {\bibfnamefont {A.}~\bibnamefont {Brosseau}}, \bibinfo {author}
  {\bibfnamefont {C.}~\bibnamefont {Lafargue}}, \bibinfo {author}
  {\bibfnamefont {V.}~\bibnamefont {Tsvirkun}}, \bibinfo {author}
  {\bibfnamefont {S.}~\bibnamefont {Lozenko}}, \bibinfo {author} {\bibfnamefont
  {S.}~\bibnamefont {Forget}}, \bibinfo {author} {\bibfnamefont
  {S.}~\bibnamefont {Ch\'enais}}, \bibinfo {author} {\bibfnamefont
  {C.}~\bibnamefont {Ulysse}}, \bibinfo {author} {\bibfnamefont
  {J.}~\bibnamefont {Zyss}}, \ and\ \bibinfo {author} {\bibfnamefont
  {M.}~\bibnamefont {Lebental}},\ }\href {\doibase 10.1103/PhysRevA.86.043817}
  {\bibfield  {journal} {\bibinfo  {journal} {Phys. Rev. A}\ }\textbf {\bibinfo
  {volume} {86}},\ \bibinfo {pages} {043817} (\bibinfo {year}
  {2012})}\BibitemShut {NoStop}%
\bibitem [{\citenamefont {Gozhyk}\ \emph {et~al.}(2015)\citenamefont {Gozhyk},
  \citenamefont {Boudreau}, \citenamefont {Haghighi}, \citenamefont {Djellali},
  \citenamefont {Forget}, \citenamefont {Ch\'enais}, \citenamefont {Ulysse},
  \citenamefont {Brosseau}, \citenamefont {Pansu}, \citenamefont {Audibert},
  \citenamefont {Gauvin}, \citenamefont {Zyss},\ and\ \citenamefont
  {Lebental}}]{Gozhyk2015}%
  \BibitemOpen
  \bibfield  {author} {\bibinfo {author} {\bibfnamefont {I.}~\bibnamefont
  {Gozhyk}}, \bibinfo {author} {\bibfnamefont {M.}~\bibnamefont {Boudreau}},
  \bibinfo {author} {\bibfnamefont {H.~R.}\ \bibnamefont {Haghighi}}, \bibinfo
  {author} {\bibfnamefont {N.}~\bibnamefont {Djellali}}, \bibinfo {author}
  {\bibfnamefont {S.}~\bibnamefont {Forget}}, \bibinfo {author} {\bibfnamefont
  {S.}~\bibnamefont {Ch\'enais}}, \bibinfo {author} {\bibfnamefont
  {C.}~\bibnamefont {Ulysse}}, \bibinfo {author} {\bibfnamefont
  {A.}~\bibnamefont {Brosseau}}, \bibinfo {author} {\bibfnamefont
  {R.}~\bibnamefont {Pansu}}, \bibinfo {author} {\bibfnamefont {J.-F.}\
  \bibnamefont {Audibert}}, \bibinfo {author} {\bibfnamefont {S.}~\bibnamefont
  {Gauvin}}, \bibinfo {author} {\bibfnamefont {J.}~\bibnamefont {Zyss}}, \ and\
  \bibinfo {author} {\bibfnamefont {M.}~\bibnamefont {Lebental}},\ }\href
  {\doibase 10.1103/PhysRevB.92.214202} {\bibfield  {journal} {\bibinfo
  {journal} {Phys. Rev. B}\ }\textbf {\bibinfo {volume} {92}},\ \bibinfo
  {pages} {214202} (\bibinfo {year} {2015})}\BibitemShut {NoStop}%
\bibitem [{\citenamefont {Braun}\ \emph {et~al.}(2000)\citenamefont {Braun},
  \citenamefont {Ihlein}, \citenamefont {Laeri}, \citenamefont {N{\"o}ckel},
  \citenamefont {Schulz-Ekloff}, \citenamefont {Sch{\"u}th}, \citenamefont
  {Vietze}, \citenamefont {Wei{\ss}},\ and\ \citenamefont
  {W{\"o}hrle}}]{Braun2000}%
  \BibitemOpen
  \bibfield  {author} {\bibinfo {author} {\bibfnamefont {I.}~\bibnamefont
  {Braun}}, \bibinfo {author} {\bibfnamefont {G.}~\bibnamefont {Ihlein}},
  \bibinfo {author} {\bibfnamefont {F.}~\bibnamefont {Laeri}}, \bibinfo
  {author} {\bibfnamefont {J.~U.}\ \bibnamefont {N{\"o}ckel}}, \bibinfo
  {author} {\bibfnamefont {G.}~\bibnamefont {Schulz-Ekloff}}, \bibinfo {author}
  {\bibfnamefont {F.}~\bibnamefont {Sch{\"u}th}}, \bibinfo {author}
  {\bibfnamefont {U.}~\bibnamefont {Vietze}}, \bibinfo {author} {\bibfnamefont
  {{\"O}.}~\bibnamefont {Wei{\ss}}}, \ and\ \bibinfo {author} {\bibfnamefont
  {D.}~\bibnamefont {W{\"o}hrle}},\ }\href {\doibase 10.1007/s003400000210}
  {\bibfield  {journal} {\bibinfo  {journal} {Appl. Phys. B}\ }\textbf
  {\bibinfo {volume} {70}},\ \bibinfo {pages} {335} (\bibinfo {year}
  {2000})}\BibitemShut {NoStop}%
\bibitem [{\citenamefont {Nobis}\ and\ \citenamefont
  {Grundmann}(2005)}]{Nobis2005}%
  \BibitemOpen
  \bibfield  {author} {\bibinfo {author} {\bibfnamefont {T.}~\bibnamefont
  {Nobis}}\ and\ \bibinfo {author} {\bibfnamefont {M.}~\bibnamefont
  {Grundmann}},\ }\href {\doibase 10.1103/PhysRevA.72.063806} {\bibfield
  {journal} {\bibinfo  {journal} {Phys. Rev. A}\ }\textbf {\bibinfo {volume}
  {72}},\ \bibinfo {pages} {063806} (\bibinfo {year} {2005})}\BibitemShut
  {NoStop}%
\bibitem [{\citenamefont {Li}\ \emph {et~al.}(2006)\citenamefont {Li},
  \citenamefont {Zhou}, \citenamefont {Zheng},\ and\ \citenamefont
  {Poon}}]{Li2006}%
  \BibitemOpen
  \bibfield  {author} {\bibinfo {author} {\bibfnamefont {C.}~\bibnamefont
  {Li}}, \bibinfo {author} {\bibfnamefont {L.}~\bibnamefont {Zhou}}, \bibinfo
  {author} {\bibfnamefont {S.}~\bibnamefont {Zheng}}, \ and\ \bibinfo {author}
  {\bibfnamefont {A.~W.}\ \bibnamefont {Poon}},\ }\href {\doibase
  10.1109/JSTQE.2006.883150} {\bibfield  {journal} {\bibinfo  {journal} {IEEE
  J. Sel. Top. Quant. Electron.}\ }\textbf {\bibinfo {volume} {12}},\ \bibinfo
  {pages} {1438} (\bibinfo {year} {2006})}\BibitemShut {NoStop}%
\bibitem [{\citenamefont {Yoon}\ \emph {et~al.}(2007)\citenamefont {Yoon},
  \citenamefont {An}, \citenamefont {Kim}, \citenamefont {Ku},\ and\
  \citenamefont {Kwon}}]{Yoon2007}%
  \BibitemOpen
  \bibfield  {author} {\bibinfo {author} {\bibfnamefont {J.}~\bibnamefont
  {Yoon}}, \bibinfo {author} {\bibfnamefont {S.-J.}\ \bibnamefont {An}},
  \bibinfo {author} {\bibfnamefont {K.}~\bibnamefont {Kim}}, \bibinfo {author}
  {\bibfnamefont {J.~K.}\ \bibnamefont {Ku}}, \ and\ \bibinfo {author}
  {\bibfnamefont {O.}~\bibnamefont {Kwon}},\ }\href {\doibase
  10.1364/AO.46.002969} {\bibfield  {journal} {\bibinfo  {journal} {Appl.
  Opt.}\ }\textbf {\bibinfo {volume} {46}},\ \bibinfo {pages} {2969} (\bibinfo
  {year} {2007})}\BibitemShut {NoStop}%
\bibitem [{\citenamefont {Yang}\ and\ \citenamefont {Huang}(2007)}]{Yang2007}%
  \BibitemOpen
  \bibfield  {author} {\bibinfo {author} {\bibfnamefont {Y.-D.}\ \bibnamefont
  {Yang}}\ and\ \bibinfo {author} {\bibfnamefont {Y.-Z.}\ \bibnamefont
  {Huang}},\ }\href {\doibase 10.1103/PhysRevA.76.023822} {\bibfield  {journal}
  {\bibinfo  {journal} {Phys. Rev. A}\ }\textbf {\bibinfo {volume} {76}},\
  \bibinfo {pages} {023822} (\bibinfo {year} {2007})}\BibitemShut {NoStop}%
\bibitem [{\citenamefont {Lafargue}\ \emph {et~al.}(2014)\citenamefont
  {Lafargue}, \citenamefont {Lebental}, \citenamefont {Grigis}, \citenamefont
  {Ulysse}, \citenamefont {Gozhyk}, \citenamefont {Djellali}, \citenamefont
  {Zyss},\ and\ \citenamefont {Bittner}}]{Lafargue2014}%
  \BibitemOpen
  \bibfield  {author} {\bibinfo {author} {\bibfnamefont {C.}~\bibnamefont
  {Lafargue}}, \bibinfo {author} {\bibfnamefont {M.}~\bibnamefont {Lebental}},
  \bibinfo {author} {\bibfnamefont {A.}~\bibnamefont {Grigis}}, \bibinfo
  {author} {\bibfnamefont {C.}~\bibnamefont {Ulysse}}, \bibinfo {author}
  {\bibfnamefont {I.}~\bibnamefont {Gozhyk}}, \bibinfo {author} {\bibfnamefont
  {N.}~\bibnamefont {Djellali}}, \bibinfo {author} {\bibfnamefont
  {J.}~\bibnamefont {Zyss}}, \ and\ \bibinfo {author} {\bibfnamefont
  {S.}~\bibnamefont {Bittner}},\ }\href {\doibase 10.1103/PhysRevE.90.052922}
  {\bibfield  {journal} {\bibinfo  {journal} {Phys. Rev. E}\ }\textbf {\bibinfo
  {volume} {90}},\ \bibinfo {pages} {052922} (\bibinfo {year}
  {2014})}\BibitemShut {NoStop}%
\bibitem [{\citenamefont {Arnaud}(1969)}]{Arnaud1969}%
  \BibitemOpen
  \bibfield  {author} {\bibinfo {author} {\bibfnamefont {J.~A.}\ \bibnamefont
  {Arnaud}},\ }\href {\doibase 10.1364/AO.8.000189} {\bibfield  {journal}
  {\bibinfo  {journal} {Appl. Opt.}\ }\textbf {\bibinfo {volume} {8}},\
  \bibinfo {pages} {189} (\bibinfo {year} {1969})}\BibitemShut {NoStop}%
\bibitem [{\citenamefont {Nixon}\ \emph {et~al.}(2013)\citenamefont {Nixon},
  \citenamefont {Redding}, \citenamefont {Friesem}, \citenamefont {Cao},\ and\
  \citenamefont {Davidson}}]{Nixon2013}%
  \BibitemOpen
  \bibfield  {author} {\bibinfo {author} {\bibfnamefont {M.}~\bibnamefont
  {Nixon}}, \bibinfo {author} {\bibfnamefont {B.}~\bibnamefont {Redding}},
  \bibinfo {author} {\bibfnamefont {A.~A.}\ \bibnamefont {Friesem}}, \bibinfo
  {author} {\bibfnamefont {H.}~\bibnamefont {Cao}}, \ and\ \bibinfo {author}
  {\bibfnamefont {N.}~\bibnamefont {Davidson}},\ }\href {\doibase
  10.1364/OL.38.003858} {\bibfield  {journal} {\bibinfo  {journal} {Opt.
  Lett.}\ }\textbf {\bibinfo {volume} {38}},\ \bibinfo {pages} {3858} (\bibinfo
  {year} {2013})}\BibitemShut {NoStop}%
\bibitem [{\citenamefont {Pascal}\ \emph {et~al.}(2017)\citenamefont {Pascal},
  \citenamefont {Bittner}, \citenamefont {Dietz}, \citenamefont {Trabattoni},
  \citenamefont {C.}, \citenamefont {Romanelli}, \citenamefont {Brunel},
  \citenamefont {Zyss},\ and\ \citenamefont {Lebental}}]{Pascal2017}%
  \BibitemOpen
  \bibfield  {author} {\bibinfo {author} {\bibfnamefont {A.}~\bibnamefont
  {Pascal}}, \bibinfo {author} {\bibfnamefont {S.}~\bibnamefont {Bittner}},
  \bibinfo {author} {\bibfnamefont {B.}~\bibnamefont {Dietz}}, \bibinfo
  {author} {\bibfnamefont {A.}~\bibnamefont {Trabattoni}}, \bibinfo {author}
  {\bibfnamefont {U.}~\bibnamefont {C.}}, \bibinfo {author} {\bibfnamefont
  {M.}~\bibnamefont {Romanelli}}, \bibinfo {author} {\bibfnamefont
  {M.}~\bibnamefont {Brunel}}, \bibinfo {author} {\bibfnamefont
  {J.}~\bibnamefont {Zyss}}, \ and\ \bibinfo {author} {\bibfnamefont
  {M.}~\bibnamefont {Lebental}},\ }\href {\doibase 10.1088/1361-6404/aa609d}
  {\bibfield  {journal} {\bibinfo  {journal} {Eur. J. Phys.}\ }\textbf
  {\bibinfo {volume} {38}},\ \bibinfo {pages} {034011} (\bibinfo {year}
  {2017})}\BibitemShut {NoStop}%
\bibitem [{\citenamefont {Jackson}(1999)}]{jackson}%
  \BibitemOpen
  \bibfield  {author} {\bibinfo {author} {\bibfnamefont {J.~D.}\ \bibnamefont
  {Jackson}},\ }\href@noop {} {\emph {\bibinfo {title} {Classical
  Electrodynamics}}}\ (\bibinfo  {publisher} {John Wiley \& Sons, Inc.},\
  \bibinfo {year} {1999})\BibitemShut {NoStop}%
\end{thebibliography}
\end{document}